# SINGLE PHOTON INTERFERENCE AND QUANTUM-CLASSICAL PHASETRANSITION WITH SPONTANEPOS SUPERPOSITION BREAKING

Vladan Panković

Department of Physics, Faculty of Sciences, 21000 Novi Sad,
Trg Dositeja Obradovića 4. , Serbia, vladan.pankovic@df.uns.ac.rs

In this work we accurately consider quantum dynamical interaction (or entanglement) between single photon and usual photographic plate as photon coordinate detector in well-known experiments of single photon interference at beam splitter (and analogous experiments). Also, we here consider final measurement or detection procedure (collapse) as a quantum-classical continuous phase transition with spontaneous (non-dynamical) unitary symmetry (superposition) breaking (effective hiding). Practically, collapse as an effective (non-absolute) phenomena can be considered as an especial case of the general formalism of spontaneous symmetry breaking (with applications in many different domains of the physics, e.g. in elasticity of rigid bodies, quantum theory of ferromagnetism, quantum theory of electro-weak interactions as well as in chaotic inflation cosmology).

## 1. Introduction

As it is well-known within standard quantum mechanical formalism [1]- [8] measurement or detection of some observable at quantum system by measurement device or detector includes two principally different processes.

First one represents strictly deterministic, unitary symmetric quantum dynamical interaction between detected quantum system and corresponding detector. It, by corresponding deterministic unitary symmetric quantum dynamical evolution, extends initial superposition of the quantum states of quantum system at final entangled quantum state as an especial superposition of the quantum states of quantum super-system (quantum system+detector). Or mentioned unitary symmetric quantum dynamical interaction entangles quantum system (precisely bi-orthogonally expands entangled quantum state over correlated eigen basis of the detected observable of quantum system and "pointer" observable of the detector) and conserves superposition. Fundamental consequence of the unitary symmetry of the quantum dynamics is that all quantum referential frames, i.e. all eigen bases of the observables (Hermitean operators in Hilbert space as basic physical space) have the same right by description of the quantum dynamics of the quantum states of unit norm as in Hilbert space as basic physical space. Simply speaking quantum dynamics does not admit existence of some absolute quantum referential frame or ”absolute observer”. It is deeply conceptually analogous with theory of relativity [7], [8]. In this sense it can be stated that role of the absolute collapse in quantum mechanics corresponds to role of the absolute space and time in classical mechanics.

Second one process represents mentioned entanglement (superposition) breaking or collapse, i.e. transition from mentioned entangled quantum state in corresponding inherently statistical mixture of the quantum states of quantum super-system (quantum system+detector). In usual phenomenological description of the detection procedure [7], [8],when detector is finally effectively described in some way “classically” (by classical mechanical dynamics), then, relative to so-described detector, quantum system is finally effectively described by inherently statistical mixture of the eigen states of detected observable. Effective “classical” description of the detector needs the following Firstly it needs that exact quantum eigen states of the “pointer” observable belong to orbital Hilbert space (where all observables represent analytical functions of the coordinate and momentum observables). Secondly it needs that all mentioned “pointer” observable exact quantum eigen states can be in well-known way [6] approximated by (extremely slowly dissipated) wave packets as effective model of the “classical particles” Thirdly, it needs [9], [10] that all mentioned wave packets are weekly interfering. Such effectively “classically” described detector effectively obtains character of an effectively ”absolute” referential frame as well as uniquely correlated quantum referential frame corresponding to eigen basis of the detected obser4vable of the quantum system. But as it is well-known it can be simply demonstrated that mentioned three necessary conditions are n0t sufficient for complete description of the effective collapse. Namely, after application of all three mentioned conditions, exact quantum entanglement, i.e. super-systemic superposition, turns out only in partially quantum (over detected quantum system) and partially “classically” (over detector) described entanglement. It is exactly different from corresponding effective statistical mixture at exact quantum level of analysis accuracy.

Since collapse cannot be exactly realized dynamically at exact quantum mechanical level of analysis accuracy it can be hypothetically supposed that mentioned effective experimental results of the final detection are absolute exact in the following sense. It can be hypothetically supposed that such absolute collapse can be realized by some non-trivial exact extension of the quantum dynamics. Namely such extended dynamics can hold additional dynamical terms that exactly and absolutely break unitary symmetry of the quantum dynamics but which stand effectively hidden at quantum mechanical level of analysis accuracy. It can be hypothetically supposed that such hidden dynamical variables belong to sub-quantum level of analysis accuracy [11] ,or to quasi-classical level of analysis accuracy [12] , or even that such hidden dynamical variables represent (thermo)dynamical decoherence as influence of the physical environment (dependence of the environment temperature, number of the physical systems in environment, etc.) at quantum system [13], [14]. (Formally, even von Neumann Abstract Ego of the human observer [1] can be considered as a meta-physical hidden variable.) But in all mentioned cases, i.e. generally, it is well-known theoretically and experimentally [15]- [17], that such hypothetical hidden variable dynamics must be necessarily super-luminal what is extremely non-plausible. Also, according to recent experimental results [18], quantum entanglement stands conserved at quantum super-systems with extremely large number of the quantum sub-systems at extremely large temperatures which sharply contradicts to decoherence or environmental hidden variables theories.

In this work we shall accurately consider quantum dynamical interaction (entanglement) between single photon and usual photographic plate as photon coordinate detector in well-known experiments of single photon interference at beam splitter (and analogous experiments) [2],[5], [7], [19],[20]. Also, we shall here consider final measurement or detection procedure (collapse) as a quantum-classical continuous phase transition with spontaneous (non-dynamical) unitary symmetry (superposition) breaking (effective hiding) [21], [22]. Practically, collapse as an effective (non-absolute) phenomena can be considered as an especial case of the general formalism of spontaneous symmetry breaking (with applications in many different domains of the physics, e.g. in elasticity of rigid bodies, quantum theory of ferromagnetism, quantum theory of electro-weak interactions as well as in chaotic inflation cosmology) [23] [26].

## 2. Quantum-classical phase transition with spontaneous superposition breaking

Consider for some quantum system, quantum mechanical dynamical state
from orbital Hilbert space. Orbital Hilbert space represents an especial sub-space of the usual Hilbert space. Over this sub-space only such observables act that represent analytical functions of the coordinate observable **x=**x or/and momentum observable **p**.

Then, as it is well-known [6] , Schrödinger exact quantum mechanical dynamical equation for quantum state can be changed by equivalent Ehrenfest exact quantum mechanical dynamical equations for average values

(1) $m\, d\langle \mathbf{x} \rangle/dt = \langle \mathbf{p} \rangle$

(2) $d\langle \mathbf{p} \rangle/dt = - \langle \partial \mathbf{V}(x)/\partial x \rangle$

Here m represents mass and **V**(x) - potential energy observable of quantum system.

Right hand of (2) can be Taylor expanded in a relatively small vicinity $\Delta x$ of $\langle x \rangle$, where $\Delta x$ represents standard deviation of x. It yields

(3) $$\begin{aligned} \langle \partial \mathbf{V}(x)/\partial x \rangle &= \partial \mathbf{V}(\langle x \rangle)/\partial \langle x \rangle + [\partial^2 \mathbf{V}(\langle x \rangle)/ \partial \langle x \rangle^2]\langle x-\langle x \rangle \rangle + \\ &\quad (1/2)\, [\partial^3 \mathbf{V}(\langle x \rangle)/ \partial \langle x \rangle^3]\langle x-\langle x \rangle \rangle^2 + \ldots \\ &= \partial \mathbf{V}(\langle x \rangle)/\partial \langle x \rangle + (1/2)\, [\partial^3 \mathbf{V}(\langle x \rangle)/ \partial \langle x \rangle^3]\, \Delta x^2 + \ldots \end{aligned}$$

In general case series (3) diverges. But series (3) can converge and first term in Taylor expansion becomes predominant for some well-known [6] quantum mechanical dynamical states called wave packets that belong to orbital Hilbert space.

Within wave-packet approximation standard deviation of any observable that acts over orbital Hilbert space (including coordinate and momentum observable) can be formally neglected in respect to absolute value of average value of mentioned observable. However as it is well-known [6] there is down limit of applicability of such approximation (including Heisenberg uncertainty relations).

Within satisfied wave packet approximation equation (1) stands practically unchanged while equation (2) turns out in the following equation

(4) $d\langle \mathbf{p} \rangle/dt = - \partial \mathbf{V}(\langle x \rangle)/\partial \langle x \rangle$

It corresponds, obviously, to Newton classical mechanical dynamics for a particle with coordinate $\langle x \rangle$ and momentum $\langle \mathbf{p} \rangle$ for classical mechanical force $-\partial \mathbf{V}(\langle x \rangle)/\partial \langle x \rangle$.

Exact quantum mechanical dynamical state $|\Psi>$ in orbital Hilbert space always exists. But only when this quantum state satisfies all mentioned approximation conditions it represents approximately a wave packet $|\Psi, WP>$ placed in small vicinity of $<x>$

It is very important that the following be observed and pointed out. During continuous unitary dynamical evolution of always existing exact quantum mechanical dynamical state $|\Psi>$ during time, cases of satisfaction of wave packet approximation conditions and wave packet $|\Psi, WP>$ existence can continuously turn out in cases of non-satisfaction of wave packet approximation conditions and wave packet $|\Psi, WP>$ non-existence (e.g. by wave packet dissipation) and vice versa. All this can be formally considered as corresponding continuous phase transitions.

Thus, exact quantum mechanical dynamics is always exactly globally stable in the orbital Hilbert's space. Also, when wave packet approximation is satisfied, quantum mechanical dynamics can be presented effectively by classical mechanical dynamics globally stable within usual coordinate space, i.e. phase space. Vice versa, when wave packet approximation is unsatisfied then exact quantum mechanical dynamics cannot be approximated classical mechanically dynamically globally stable in the usual coordinate space, i.e. phase space. Finally, transition from the situation when wave packet approximation is satisfied in situation when wave packet approximation is unsatisfied (and vice versa) can be presented as a typical continuous phase transition.

Consider, now, for some quantum system, two quantum states $|\Psi_1>$ and $|\Psi_2>$ from orbital Hilbert space so that quantum dynamical state of mentioned quantum system represents some superposition of mentioned two quantum states. As it has been previously discussed mentioned exactly existing superposition stands always conserved by mentioned quantum dynamical evolution during time.

Suppose that both mentioned quantum states $|\Psi_1>$ and $|\Psi_2>$ satisfy wave packet approximation so that wave packets $|\Psi_1, WP>$ and $|\Psi_2, WP>$ with corresponding coordinate average values and standard deviations $<x_1>$ and $\Delta x_1 = \Delta x$ and $<x_2>$ and $\Delta x_2 = \Delta x_1 = \Delta x$ simultaneously exist. However, it can be simply proved that when mentioned two wave packets weakly interfere then superposition of mentioned wave packets does not represent any wave packet. Consider for simplicity superposition when absolute vale of superposition coefficients are proportional or practically almost equivalent. Namely then absolute value of the coordinate average value in superposition of wave packets equals practically $|(<x_1> + <x_2>)/2|$ while coordinate standard deviation in superposition of wave packets equals practically $|(<x_1> - <x_2>)/2|$. Further we can consider $|(<x_1> - <x_2>)|$ as order parameter with critical value $\Delta x$ so that for order parameter smaller than critical value two wave packets overlap and interfere strongly, while in opposite case, i.e. for order parameter larger than critical value two wave packets weakly interfere. In last case of weakly interfering two wave packets superposition of mentioned wave packets does not satisfy all wave packet conditions (distribution of the quantum system coordinate nearly $|(<x_1> + <x_2>)/2|$ can be neglected, Taylor expansion of Ehrenfest average value of the quantum dynamical equation becomes divergent, etc.). For this reason exactly existing quantum superposition of two or more weakly interfering wave packets, approximately classically is globally non-stable and it cannot be globally presented as any wave packet.

However, mentioned superposition effectively can be locally classically presented either by one ("left") or by other ("right") wave packet in corresponding small coordinate domains in which mentioned wave packets are locally classically well-defined. This situation is conceptually completely analogous to situation by spontaneous gauge symmetry breaking in electro-weak interactions [23]- [25].

As it is well-known theory of electro-weak interaction must be exactly gauge symmetric since only in this case mentioned theory admits renormalization. For this reason boson masses cannot be obtained by introduction of additional dynamical terms which exactly break gauge symmetry. However boson masses can be obtained by spontaneous (non-dynamical) breaking (effective hiding) of the gauge symmetry. Simply
speaking, exact dynamical equations of the quantum field theory for electro-weak interactions must be solved iteratively approximately, by perturbation theory (which has character of a Taylor expansion).Within false vacuum quantum field point perturbation theory globally diverges, while within a circular ring of real vacuum quantum field points perturbation theory locally converges. For this reason from perturbation theory approximate level of accuracy, dynamical characteristics of the false vacuum are dynamically globally non-stable and non-describable, while dynamical characteristics of the real vacuums are dynamically locally stable and describable. Formally and
effectively false vacuum can be changed by one of many equally probable real vacuums when circular symmetry of the real vacuums as well as gauge symmetry becomes spontaneously (non-dynamically) broken (effectively hidden). Mentioned change and irreduciblely statistical choice characterize in fact perturbation theory and have not absolute character (mentioned choice of some real vacuum is not governed by some small additional dynamical term).

All this admits formulation of the general formalism of the spontaneous (non-dynamical) symmetry breaking (effective hiding) in the following way. Simply speaking by dynamical symmetry breaking in some of many different physical theories dynamical symmetry at some level of analysis accuracy is broken by dynamical terms effectively hidden at this accuracy level but really existing at more exact accuracy level. It can be added that dynamical symmetry breaking represents physical content for many usual gambling games, e.g. at play at dice etc. . It can imply a seeming impression that dynamical symmetry breaking represents unique physical corner stone for mathematical theory of probability. By spontaneous symmetry breaking in some of many different physical theories (e.g. in elasticity of rigid bodies, quantum theory of ferromagnetism, quantum theory of electro-weak interactions as well as in chaotic inflation cosmology [23]- [26]) dynamical symmetry and dynamical state really exist at exact level of analysis accuracy. However, dynamical state can be expressed only approximately, by some iterative, e.g. Taylor expansion series including perturbation theories. Mentioned Taylor series can converge not in all domains, i.e. generally or globally but only in some different domains, i.e. particularly or locally. For this reason exactly existing symmetric dynamical state becomes effectively hidden or effectively dynamically non-stable for approximate iterative level of analysis accuracy. It can be locally projected or changed by some particular dynamical state that is locally effectively dynamically stable within corresponding local domain of Taylor series convergence. Simultaneously mentioned change implies effective hiding of the symmetry. Obviously mentioned projection of effectively dynamically non-stable state in effectively dynamically stable state is effectively irreversible and it cannot be completely described neither by exact nor by effective dynamics. However, mentioned projection can be effectively considered as inherently statistical or probabilistic event. Of course, after realization of mentioned event effective dynamical state evolves strictly deterministically according to effective dynamics. All this is metaphorically and epistemologically analogous to well-known story about competition between three tailors in the same street in some town. First tailor stated that he is best tailor in the town. Second tailor stated that he is the best tailor in the world. But the third tailor, winner, stated only that he is the best tailor in this street.

According to general formalism of the spontaneous symmetry breaking it can be supposed [21], [22] that superposition of weakly interfering wave packets, within classical mechanical approximation only, effectively self-collapses or turns out spontaneously (non-dynamically) in statistical mixture of mentioned wave packets. On the other hand, at exact quantum level of analysis accuracy, there is no any breaking of the superposition of corresponding wave packets. Self-collapsed state, at approximate level of analysis accuracy, evolves further classical mechanical dynamically deterministically and it cannot here return in initial approximately classical dynamically globally non-stable state. On the other hand at exact quantum level of analysis accuracy where entangled quantum state exists self-collapse has been never realized and exact quantum phenomena are reversible. It simply explain previously discussed principal difference between reversible exact unitary symmetric quantum dynamical evolution and effectively classically irreversible collapse [1]. All this admits metaphorical statement that (self)collapse represents cut of previously mentioned third taylor.

Since any superposition of weakly interfering wave packets effectively self-collapses it implies that basis of all weakly interfering wave packets has within mentioned approximation especial, effectively dominant role so that it can be formally considered as “absolute” referential frame and self-collapse as “absolute” collapse. In this way weakly interfering wave packets can be effectively considered as the eigen states of the “absolute” “pointer” observable.

Self-collapse as continuous phase transition with spontaneous superposition breaking can be presented in the following way. Suppose that initially all wave packets strongly interfere. Suppose that, during quantum dynamical evolution, distance between any two wave packet centers (average coordinate values) as order parameters increase while coordinate deviation of any wave packet stands practically the same (wave packets non-dissipation condition). Then during time any two wave packets interfere weaker and weaker and finally all wave packets become mutually weakly interfering or practically non-interfering when self-collapse appears.

So, self-collapse as the spontaneous superposition breaking can be realized only at quantum states from orbital Hilbert space that represent superposition of weakly interfering wave packets during time intervals within which all wave packet approximation conditions are satisfied. It means that self-collapse can be in principle realized at any physical system, microscopic or macroscopic. However for microscopic systems with relatively small masses wave packets dissipate very quickly while for macroscopic systems with relatively large masses wave packets dissipate very slowly. For this reason classical mechanical description of microscopic system or quantum mechanical description of macroscopic systems are technically extremely hard for realization even if both, first effectively and second exactly, are possible in principle.

Quantum states that do not belong to orbital Hilbert space of some quantum system , “object”, cannot realize self-collapse at all. Also, when conditions for self-collapse are not satisfied, “object” quantum state from orbital Hilbert space cannot realize self-collapse. However, “object” can quantum dynamically interact with appropriate other quantum system, “detector” at which self-collapse can effectively occur. Suppose that by mentioned quantum dynamical interaction entanglement between “detector” and “object” is exactly realized so that exactly existing entangled quantum state is bi-orthogonally expanded over eigen basis of “detector” “pointer” observable and eigen basis of some “object” observable. In this case when at “detector” self-collapse effectively appears, i.e. when some eigen state of the “detector” “pointer” observable appears with corresponding statistical weight, then, via entanglement, at “object” correlated eigen state of mentioned observable appears with (conditional) certainty. In this sense we can speak about relative collapse at "object" as effectively pure quantum phenomena.

It represents general theory of the detection procedure (collapse) as a quantum-classical continuous phase transition with spontaneous (non-dynamical) unitary symmetry (superposition) breaking (effective hiding).

### 3. Single photon interference detected by photographic plate

Consider now firstly usual experiment of single photon interference at beam splitter without photon trajectories detection (presented symbolically at figure 1.). Here source S emits single photon (wavefront) with well-defined momentum and corresponding wave length that propagates toward beam splitter BS. "Reflected" photon (redirected by first total mirror) propagates toward photographic plate PP that represents photon coordinate detector, while "transmitted" photon (redirected by second total mirror ) propagates toward PP too. Then, immediately before detection that PP realizes at photon, photon is described by the following quantum superposition state

(5) $|ph> = c_1 |R> + c_2 |T>$

Here $|R>$ and $|T>$ represent "reflected" and "transmitted" quantum state (trajectory) while $c_1$ and $c_2$ represent superposition coefficients (which includes especial case characteristic for usual experiment where mentioned superposition coefficients equal $i\, 2^{-1/2}$ and $2^{-1/2}$ respectively). Final detection of single photon (coordinate), i.e. absorption of single photon can occur in some "point", i.e. elementary segment (part) of PP.

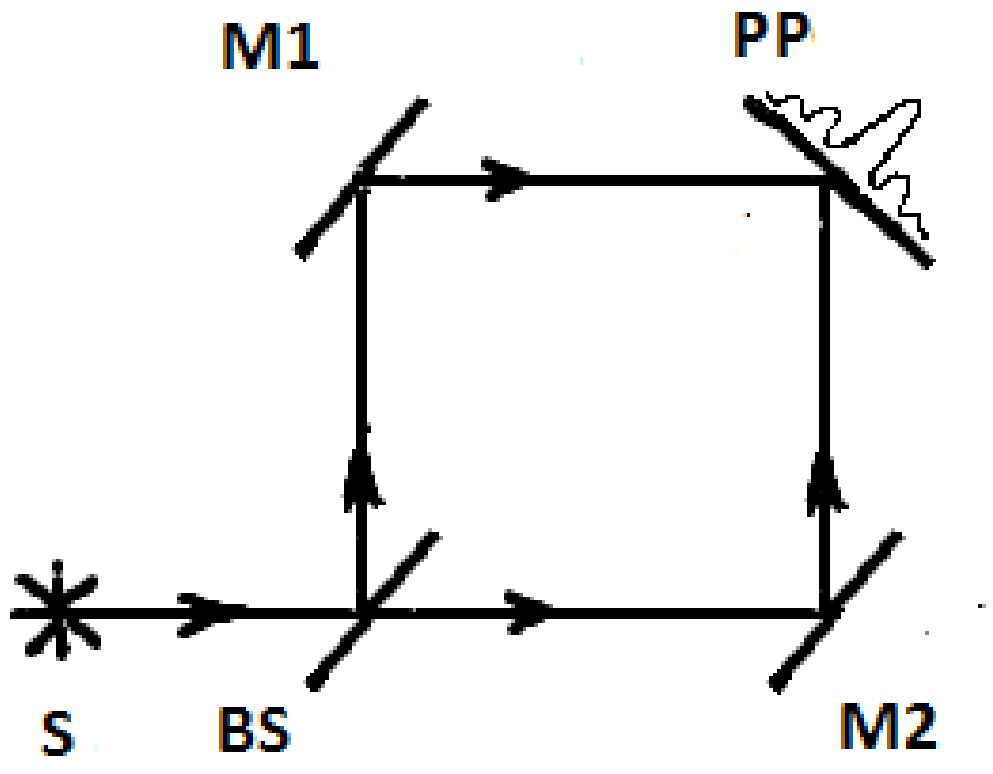

Figure 1.

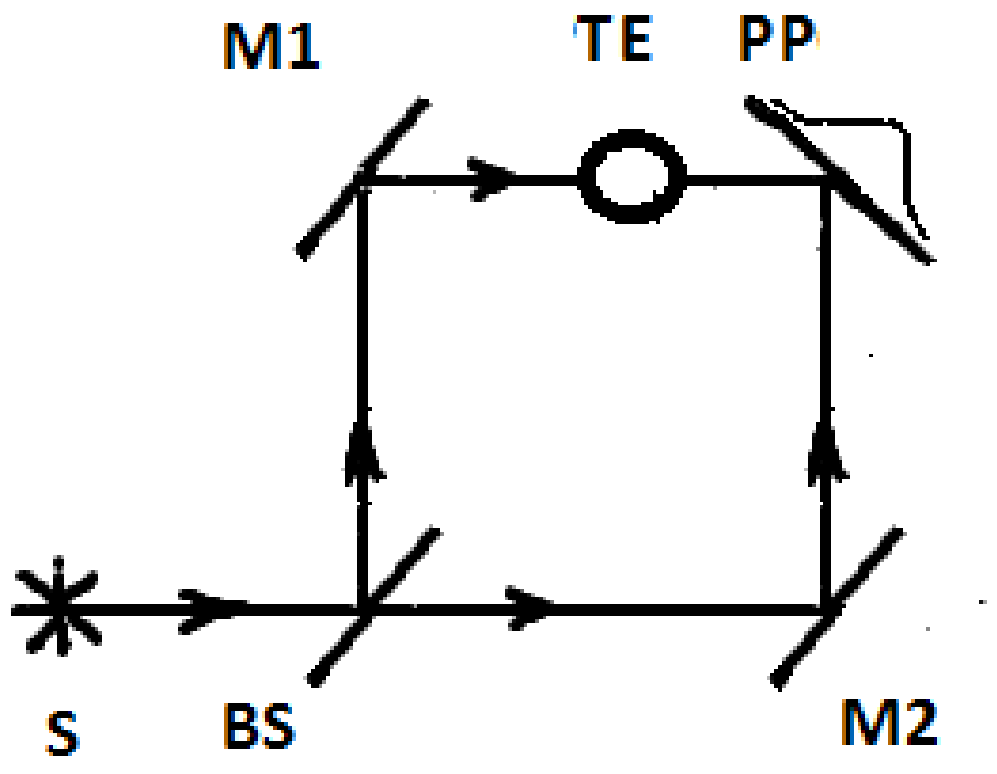

Figure 2.

We shall now accurately consider how PP quantum dynamically interacts with single photon described by (5).

Suppose that PP (without any diminishing of the basic conclusions) can be presented by the following simplified model.

Suppose that PP represents one-dimensional physical system with characteristic space direction q. Also suppose that PP consists of many identical, quantum dynamically non-interacting elementary segments EPP1, EPP2, … , EPPn placed in series along q direction. Suppose that in any elementary segment of PP there is a number of photo
-sensitive micro-elements, photo-sensitive micro-crystals of the same type of silver halide which are practically homogeneously distributed. Finally suppose that EPPn has a small length $\Delta q$ and that EPPn occupies coordinate interval
$I_n(q) = ((n-1)\Delta q , n\Delta q )$ with centers in $q_n = (n-1/2)\Delta q$ for n=1,2, … .

Define |n,0 > as quantum state of EPPn without absorption of any photon for
n =1,2, …..

Define |n,1 > as quantum state of EPPn after absorption of single photon for
n =1,2, ….. . By absorption of single photon photo-sensitive micro-element changes principally its characteristics. For example, as it is well-known, absorption of single photon by silver halide micro-crystal corresponds to transition of corresponding single electron from valence band (where mentioned electron is localized nearly some atom) in conduction band (where mentioned electron is delocalized and can move through crystal). Such delocalized electron is attracted toward an electron trap (crystalline defect etc.) at sensitivity speck at crystal surface. Here finally electron is recombined with some interstitial silver ion in neutral silver atom (in common with corresponding energy, t.e. photon emission). It can be observed that initial quantum state of electron localized nearly some atom in crystal valence band and final quantum state of electron localized in silver atom in sensitivity speck extremely weakly interfere. For this reason |n,0 > and |n,1 > extremely weakly interfere too for n=1,2, … .

Define “zero” eigen state of PP “pointer” observable

(6) $|PP,0 > = |1,0 >|2,0 > \dots \quad | n,0 >$

Define n-th eigen state of PP “pointer” observable

(7) $|PP,n > = |1,0 >|2,0 > \dots|n-1,0 >| n,1 >| n+1,0 > \dots$

for n =1,2, …..

Finally suppose that PP before quantum dynamical interaction with photon is described by quantum state (6) while simultaneously photon is described by quantum state (5). It implies that quantum super-system photon+PP before quantum dynamical interaction between photon and PP is described by non-entangled quantum state
|ph >|PP,0 >.

Define |ph,q,n > as the quantum state of photon when q coordinate of photon belongs to coordinate interval $I_n(q)$ for n=1,2, … .

Then quantum state of photon |ph >, immediately before quantum dynamical interaction between photon and PP, can be presented by the following expression

(8) $|ph > = \sum_n < ph,q,n |ph > |ph,q,n >$

that generally represents superposition of quantum states |ph,q,n > with superposition coefficients < ph,q,n |ph > for n =1,2, …. . According to introduced suppositions expression (8) can be presented in the following concrete form

(9) $|ph> = \sum_n <ph,q,n|ph>|ph,q,n>$

Define unitary symmetric evolution operator **U** corresponding to quantum dynamical interaction between photon and PP by expression

(10) $\mathbf{U}|ph,q,n>|PP,0> = |ph,q,n>|PP,n>$

for n =1,2, …..

Then, immediately after quantum dynamical interaction between photon and PP non-entangled quantum state |ph >|PP,0 > turns out strictly deterministically in the following entangled quantum state

(11) $|ph+PP> = \mathbf{U}|ph>|PP,0> = \sum_n <ph,q,n|ph>|ph,q,n>|PP,n>$

It can be observed that in EPPn initially captured electron in valence band, i.e. nearly some atom can be approximately described by a wave packet (initial wave packet) |n,0,WP > without momentum average value, with coordinate average value corresponding to center of the domain of initial localization and wave-packet width corresponding to linear dimension of the domain of initial localization for n=1,2, … . Also in EPPn finally captured electron in neutral silver atom can be approximately described by a wave packet (final wave packet) |n,1,WP > without momentum average value, with coordinate average value corresponding to center of the domain of final localization and wave packet width corresponding to linear dimension of the domain of final localization for n=1,2, … . Obviously, in EPPn mentioned electron initial and final wave packets extremely weakly interfere for n=1,2, … . Then, in simple analogy with (6), (7), |PP,n,WP > as wave packet approximation of |PP,n > can be defined for n=0,1,2, … .Obviously wave packet approximations of all eigen states of PP "pointer" observable mutually extremely weakly interfere.

In mentioned partial wave-packet approximation entangled quantum state (11) turns out in the following entangled quantum state

(12) $\sum_n <ph,q,n|ph>|ph,q,n>|PP,n,WP>$

According to previous discussions it is not hard to see that here all conditions for appearance of the self-collapse at PP and relative3 collapse at photon aresatisfied so that in partial wave packet approximation entangled quantum state (12) turns out spontaneously (non-dynamically) in inherently statistical mixture of the states
|ph,q,n >|PP,n,WP > with statistical weights

(13) $|<ph,q,n|ph>|^2$ …

or, according to (5) ,

(14) $|<ph,q,n|(c_1|R> + c_2|T>)|^2 = |c_1<ph,q,n|R> + c_2<ph,q,n|T>|^2$

for n =1,2, … which of course corresponds to effective interference patterns appearance in full agreement with real experimental facts..

However, quantum mechanically exactly photon+PP (if we neglect further extension of entanglement via photon emitted by neutral Ag atom constitution) is exactly described by entangled quantum state (11).without any self or relative collapse appearance.

Consider now the following, second experiment of single photon interference at beam splitter (presented symbolically at figure 2. ) which, in respect to previous experiment presented, holds an additional quantum system, called symbolically photon "trajectory entangler" TE, placed at "reflected" photon trajectory, between first total mirror and PP.

It can be supposed that TE represents any quantum system (simple quantum system or quantum super-system) which can be quantum mechanically generally described in the following way. Suppose that |TE,0 >, |TE,R > and |IE,T > represent corresponding eigen states of some characteristic TE observable. Suppose that TE is initially, i.e. before quantum dynamical interaction with photon, described by |TE,0 >. It implies that, immediately before quantum dynamical interaction between TE and photon, quantum super-system TE+photon is described by non-entangled quantum state |TE,0 >|ph >. Suppose that quantum dynamical interaction between TE and photon is described by unitary evolution operator **U(**TE+ph**)** defined by the following expressions

(15) $\mathbf{U}(TE+ph)\, |TE,0\rangle |R\rangle = |TE,R\rangle |R\rangle$

-

(16) $\mathbf{U}(TE+ph)\, |TE,0\rangle |T\rangle = |TE,T\rangle |T\rangle$ .

In this way photon "reflected" trajectory becomes correlated with |TE,R > and photon "transmitted" trajectory - with |TE,T >.Then, immediately after quantum dynamical interaction between TE and photon TE+photon turns out strictly deterministically from non-entangled quantum state |TE,0 >|ph > in entangled quantum state

(17) $|TE+ph\rangle = \mathbf{U}(TE+ph)\, |TE,0\rangle|ph\rangle = c_1|TE,R\rangle |R\rangle + c_2 |TE,T\rangle|T\rangle$

(We can simply demonstrate that such dynamical interaction without significant perturbation of the photon trajectories is possible in principle. Suppose that **S** emits single photon by quantum jump of some atom from excited state |E > in ground state
|G >. Suppose now that TE consists of two distant atoms of the same type as the atoms in S, where first (closer to mirror) is initially in excited state while second (closer to PP) is initially in ground state. In this way TE is initially in quantum state |E > |G > . Obviously TE can immediately quantum dynamically interact only with "reflected" photon and this interaction can be simply described by stimulated emission of a new "copy" photon by first atom and absorption of this "copy" photon by second atom without any decoherence of the first photon "reflected" state. Simultaneously, after mentioned interaction, TE becomes described by quantum state |G > |E > . On the other hand, there is no any immediate quantum, dynamical interaction between TE and "transmitted" photon and non-interacting TE and "transmitted" photon stand in non-perturbed initial quantum states. So, here |TE,0 > = |TE,T > = |E > |G > and |TE,R > = G > |E > .It can be added that conceptually similar situation appears in real experiments of the interference of large molecules at diaphragm with two slits by trajectory detection by collisions with small molecules [19]. Here such low intensive collisions are possible, non-sufficient for significant

perturbation of the large molecule trajectories but sufficient for realization of the entanglement between large and small molecule.)

According to previous discussions and since (17) quantum mechanically exactly represents entangled quantum state (but not statistical mixture of corresponding states or trajectories) TE quantum mechanically exactly does not represent photon trajectories detectorthat generates absolute collapse.

Suppose that after previous mentioned quantum dynamical interaction between TE and photon later only photon immediately quantum dynamically interacts with PP while between TE and PP there is no any immediate quantum dynamical interaction.

So quantum mechanically exaxtly PP before quantum dynamical interaction with TE+photon is described by quantum state (6) while simultaneously TE+photon is described by entangled state (17). It implies that TE+photon+PP before quantum dynamical interaction between TE+photon and PP is described by partially entangled quantum state $|TE+ph>|PP,0>$.

As it is not hard to see according to previous suppositions quantum dynamical interaction between TE+photon and **PP** can be presented by the following unitary evolution operator **1** x **U** (for **U** that satisfies (10) ). Then initial (before mentioned interaction) partially entangled quantum state $|TE+ph>|PP,0>$ quantum dynamically deterministically evolve in the following final (after mentioned interaction) quantum state

(18) $(\mathbf{1} \times \mathbf{U})|TE+ph>|PP,0> =$
$= c_1|TE,R>\mathbf{U}|R>|PP,0> + c_2|TE,T>\mathbf{U}|T>|PP,0> =$
$= \sum_n (c_1|IE,R><ph,q,n|R> + c_2|TE,T>|<ph,q,n|T>)|ph,q,n>|PP,n>$
.

Quantum state (18) can be further transformed in the following, correctly three-orthogonally expanded, entangled quantum state

(19) $\sum_n a(n)|TEa(n)>|ph,q,n>|PP,n>$

where

(20) $a(n) = (|c_1|^2|<ph,q,n|R>|^2 + |c_2|^2|<ph,q,n|T>|^2)^{1/2}$

represents a superposition coefficient and where

(21) $|TEa(n)> =$.
$(c_1|TE,R><ph,q,n|R> + c_2|TE,T><ph,q,n|ph,T>)/a(n)$

represents a normalized quantum state of TE for n =1,2, … . It is very important to be pointed out that quantum state (21) represents a quantum superposition (but not statistical mixture) of quantum states $|TE,R>$ and $|TE,T>$ with corresponding superposition coefficients that both depend of the photon coordinate distribution for both photon trajectories.

Entangled quantum state (19) can be further partially wave-packet approximated by entangled quantum state

(22) $\sum_n a(n)|TEa(n)>|ph,q,n>|PP,n,WP>$

According to previous discussions it is not hard to see that here all conditions for appearance of the self-collapse at PP and relative3 collapse at TE+photon are satisfied so that in partial wave packet approximation entangled quantum state (22) turns out spontaneously (non-dynamically) in inherently statistical mixture of the states
|TEa(n) >|ph,q,n > |PP,n,WP > with statistical weights

(23) $|a(n)|^2 = |c_1|^2 |< ph,q,n |R >|^2 + |c_2|^2 |<ph,q,n |T >|^2$

for n =1,2, … and it of course corresponds to effective interference patterns absence in full agreement with real experimental facts. Really analogous situation we have in real experiments of mentioned low intensive collisions between large and small molecules [19] when interference patterns effectively disappear even if small mpčecules represent large molecule trajectories entanglers nut not detectors..

Here it is extremely important to be pointed out that statistical distribution (23),as it has been previously discussed, is not caused by absolute collapse realized by previous interaction between TE and photon. It is result of the collapse as the spontaneous unitary symmetry, i.e. superposition breaking without breaking of the unit norm of quantum states.

However, quantum mechanically exactly TE+photon+PP (if we neglect further extension of entanglement via photon emited by neutral Ag atom constitution) is exactly described by entangled quantum state (19).without any self or relative collapse appearance

All this simply implies that there is no any need for absolute collapse. As it is not hard to see, in case when "entangler" quantum states satisfy mentioned additional approximation conditions for spontaneous superposition breaking "entangler" turns out in "detector". But even in this case"entangler"-"detector" realizes initially the collapse only effectively and this detection is quite compatible with PP final detection. In mentioned experiment, resuulta of PP final detection are effectively independent in case when "entangler"-"detector" is presented exactly as "entangler" ("actor") or effectively as "detector" ("spectzator"). It implies that here, i.e. in experiment presented symbolically at figure 2. as well as in its relized analogies [7], entanglement between detected quantum system and "detector" (Schrödinger cat effect is realized). However, in all mentioned situations TE+photon+PP is finally exactly quantum mechanically described exclusively by entangled quantum state (19) ("however it unitary evolve", we can paraphrase Galileo Galilei).

So all exact quantum phenomena can be exactly and completely described by strictly deterministic unitary symmetric (that conserves superposition) quantum dynamical evolution (without any super-luminal phenomena) of quantum states (of the unit norm) in Hilbert space as basic physical space. In this way quantum mechanics is consistently founded. Detection or measurement can be consistently considered as a quantum-classical phase transition with spontaneous (non-dynamical) unitary symmetry (superposition) breaking (effective hiding).

## 4. Conclusion

In conclusion we can repeat and point out the following. In this work we accurately consider quantum dynamical interaction (or entanglement) between single photon and usual photographic plate as photon coordinate detector in well-known experiments of single photon interference at beam splitter (and analogous experiments). Also, we here consider final measurement or detection procedure (collapse) as a quantum-classical continuous phase transition with spontaneous (non-dynamical) unitary symmetry (superposition) breaking (effective hiding). Practically, collapse as an effective (non-absolute) phenomena can be considered as an especial case of the general formalism of spontaneous symmetry breaking (with applications in many different domains of the physics, e.g. in elasticity of rigid bodies, quantum theory of ferromagnetism, quantum theory of electro-weak interactions as well as in chaotic inflation cosmology).